\documentclass[seceq]{ptptex}

 \usepackage{latexsym,bm,amssymb,amsfonts}
  \usepackage{epsfig,graphics,graphicx}

  \newcommand{\beq}{\begin{eqnarray}}
  \newcommand{\eeq}{\end{eqnarray}}
   \newcommand{\rme}{{\rm e}}
  
   \newcommand{\rmd}{{\rm d}}

    \newcommand{\Lam}{\Lambda_{{\rm QCD}}}
   \long\def\comment#1{ }

\title{Partons and jets at strong coupling from AdS/CFT}

\author{Edmond \textsc{Iancu}%
}

\inst{Institut de Physique Th\'eorique de Saclay,
 F-91191 Gif-sur-Yvette, France}

\abst{Calculations using the AdS/CFT correspondence can be used to unveil the
short--distance structure of a strongly coupled plasma, as it would be
seen by a `hard probe'. The results of these calculations admit a natural
physical interpretation in terms of parton evolution in the plasma: via
successive branchings, essentially all partons cascade down to very small
values of the longitudinal momentum fraction $x$ and to transverse momenta
smaller than the saturation momentum $Q_s\sim T/x$. This scale $Q_s$ controls
the energy loss and the transverse momentum broadening of an energetic jet
propagating through the plasma. This picture has some striking
consequences, like the absence of jets in electron--proton annihilation
at strong coupling, of the absence of particle production at forward and
backward rapidities in hadron--hadron collisions, which look very different
from the corresponding predictions of perturbative QCD and also from the known
experimental situation.}

\begin{document}

\maketitle

\section{Introduction}

One of the most interesting suggestions emerging from the experimental
results at RHIC is that the deconfined, `quark--gluon', matter produced
in the early stages of an ultrarelativistic nucleus--nucleus collision
might be strongly interacting. This observation motivated a multitude of
applications of the AdS/CFT correspondence to problems involving a
strongly--coupled gauge plasma at finite temperature and/or finite quark
density. While early applications have focused on the long--range and
large--time properties of the plasma, so like hydrodynamics, more recent
studies have been also concerned with the response of the plasma to a
`hard probe' --- an energetic `quark' or `current' which probes the
structure of the plasma on space--time scales much shorter than the
characteristic thermal scale $1/T$ (with $T$ being the temperature).
Although the relevance of such applications to actual hard probes in QCD
is not so clear (since, by asymptotic freedom, QCD should be weakly
coupled on such short space--time separations), the results that have
been obtained in this way are conceptually interesting, in that they shed
light on a new physical regime --- that of a gauge theory with strong
interactions --- which for long time precluded all first--principles
theoretical investigations other than lattice gauge theory.

From the experience with QCD one knows that the simplest hard probe is an
electromagnetic current. In deep inelastic scattering (DIS), the exchange
of a highly virtual space--like photon between a lepton and a hadron acts
as a probe of the hadron parton structure on the resolution scales set by
the process kinematics: if $Q^2$ is (minus) the photon virtuality and $s$
is the invariant photon--hadron energy squared, then the photon couples
to quark excitations having transverse momenta $k_\perp\lesssim Q$ and a
longitudinal momentum fraction $x\sim Q^2/s$. Also, the partonic
fluctuation of a space--like current can mimic a quark--antiquark
`meson', which is nearly on--shell in a frame in which the current has a
high energy. Furthermore, the decay of the time--like photon produced in
electron--positron annihilation is the simplest device to produce and
study hadronic jets in QCD. Thus, by studying the propagation of an
energetic current through the plasma one has access to quantities like
the plasma parton distributions, the meson screening length, or the jet
energy loss.

The AdS/CFT correspondence allows one to generalize such studies to
strong coupling, at least for the case where QCD is replaced by the
${\mathcal N}=4$ supersymmetric Yang--Mills (SYM) theory.  (For recent
reviews and more references see Ref. \cite{ADSREV}.)  This theory is
conformally symmetric (it has a vanishing $\beta$ function) and thus it
differs from QCD in several important aspects: there is no confinement,
no asymptotic freedom, no intrinsic mass scale, and no fermions in the
fundamental representation of the color group SU$(N_c)$ --- rather, all
the fields appearing in the respective Lagrangian (gluons, fermions and
scalars) lie in the adjoint representation. But such differences might
not be essential for a study of the finite--temperature phase, where
confinement and the running of the coupling are presumably less important
even in QCD
--- at least, within the limited temperature interval $T=(2\div 5)T_c$,
which is the range to be explored by the heavy--ion experiments at RHIC
and LHC. (Here, $T_c\sim 200\,{\rm MeV}\sim\Lam$ is the critical
temperature for deconfinement.)

Within the ${\mathcal N}=4$ SYM theory, the role of the electromagnetic
current is played by the `${\mathcal R}$--current' --- a conserved
Abelian current whose charge is carried by (adjoint) fermion and scalar
fields. Thus, DIS at strong coupling can be formulated as the scattering
between this ${\mathcal R}$--current and some appropriate `hadronic'
target. The first such studies \cite{Polchinski,HIM1} have addressed the
zero--temperature problem, where the target was a `dilaton' --- a
massless string state `dual' to a gauge--theory `hadron', whose existence
requires the introduction of an infrared cutoff $\Lambda$ to break down
conformal symmetry. These studies led to an interesting picture for the
partonic structure at strong coupling (see also the contribution by Y.
Hatta in this volume \cite{Hatta}): through successive branchings, all
partons end up by `falling' below the `saturation line', i.e., they
occupy --- with occupation numbers of order one --- the phase--space at
transverse momenta below the saturation scale $Q_s(x)$, which itself
rises rapidly with $1/x$. Such a rapid increase, which goes like
$Q_s^2(x)\sim 1/x$ and hence is much faster than for the corresponding
scale in perturbative QCD \cite{CGC}\,, comes about because the
high--energy scattering at strong coupling is governed by a spin $j\simeq
2$ singularity (corresponding to graviton exchange in the dual string
theory), rather than the usual $j\simeq 1$ singularity associated with
gluon exchange at weak coupling.

In Refs. \cite{HIM2,HIM3} these studies and the corresponding partonic
picture have been extended to a finite--temperature ${\mathcal N}=4$ SYM
plasma and also to the case of a time--like current (the strong--coupling
analog of $e^+e^-$ annihilation). Note that this finite--$T$ case is
conceptually clearer than the zero--temperature one, in that it does not
require any `deformation' of the gauge theory, so like an IR cutoff. It
is also technically simpler, in that the calculations can be performed in
the strong 't Hooft coupling limit $\lambda\equiv g^2N_c\to \infty$ at
fixed $g^2\ll 1$ (meaning $N_c\to\infty$). This is so, roughly speaking,
since the large number of degrees of freedom in the plasma, $\sim N_c^2$
per unit volume, compensates for the $1/N_c^2$ suppression of the
individual scattering amplitudes; hence, a strong--scattering situation
can be achieved even in the strict large--$N_c$ limit. The results of
Refs. \cite{HIM2,HIM3} will be briefly described in what follows.

\section{Current--current correlator from AdS/CFT}

The strong coupling limit $\lambda\to \infty$ in the ${\mathcal N}=4$ SYM
gauge theory corresponds to the semiclassical, `supergravity',
approximation in the dual string theory, which lives in a
ten--dimensional curved space--time with metric $AdS_5\times S^5$. Our 4D
Minkowski space corresponds to the boundary of $AdS_5$ at $r\to\infty$.
($r$ is the radial dimension of $AdS_5$.) The finite--temperature gauge
plasma is `dual' to a black hole (BH) in $AdS_5$ which is homogeneous in
the four physical dimensions but has an horizon in the fifth, `radial',
dimension, located at $r_0=\pi R^2 T$. ($R$ is the curvature radius of
$AdS_5$.) The $AdS_5$ BH metric reads :
  \beq
 \rmd s^2
 =\frac{r^2}{R^2}(-f(r)\rmd t^2+\rmd \bm{x}^2)+\frac{R^2}{r^2f(r)}
 \rmd r^2\,,\qquad f(r)\,=\,1-\frac{r_0^4}{r^4}\,. \label{met2} \eeq
The ${\mathcal R}$--current $J_\mu$ acts as a perturbation on the
Minkowski boundary, that we choose as a plane wave propagating in the $z$
direction in the plasma rest frame: $J_\mu(x)\,\propto \,\rme^{-i\omega
t+iq z}$. This perturbation induces a vector field
$A_\mu(t,\bm{x},r)=\rme^{-i\omega t+iq z}A_\mu(r)$ which propagates
inside the bulk of $AdS_5$ according to the curved--space Maxwell
equations. (We work in the `radial' gauge $A_r=0$.) The interaction
between the ${\mathcal R}$--current $J_\mu$ and the plasma is then
described as the gravitational interaction between the Maxwell wave
$A_\mu$ and the BH, as encoded in the $AdS_5$ BH geometry. The
fundamental object to be computed is the retarded current--current
correlator,
   \beq
 \Pi_{\mu\nu}(q)\,\equiv\,i\int \rmd^4x\,\rme^{-iq\cdot x}\,\theta(x_0)\,
 \langle [J_\mu(x), J_\nu(0)]\rangle_T\, \label{Rdef}  \eeq
(with $q^\mu=(\omega,0,0,q)$), whose imaginary part determines the total
cross--section for the interactions of the current --- i.e., the plasma
structure functions in the {\em space--like} case $\omega^2 -q^2 < 0$
(`deep inelastic scattering') and the rate for the current decay into
`jets' in the {\em time--like} case $\omega^2 -q^2 >0$ (`$e^+e^-$
annihilation'). The imaginary part arises in the classical gravity
calculation via the condition that the wave $A_\mu$ has no reflected
component returning from the horizon. Physically, this means that the
wave (current) can be absorbed by the black hole (the plasma), but not
also regenerated by the latter. The classical solution $A_\mu(r)$ is
fully determined by this `no--reflected--wave' condition near the horizon
together with the condition that the fields take some prescribed values
at the Minkowsky boundary: $A_\mu(r\to\infty)\,=\,A^{(0)}_\mu$. The
current--current correlator is finally obtained as
 \beq
 \Pi_{\mu\nu}(q)\,=\,\frac{\partial^2 \mathcal{S}_{\rm cl}}
{\partial A_\mu^{(0)}
 \partial A_\nu^{(0)}}\,,\eeq
where $\mathcal{S}_{\rm cl}$ denotes the classical action density (the
Maxwell action evaluated on the classical solution), and is bilinear in
the boundary fields $A^{(0)}_\mu$.

\section{Parton branching in the vacuum: no jets at strong coupling}

At a first sight, the above prescription for computing $\Pi_{\mu\nu}$
might look like a `black box', but this is not really so:  The physical
interpretation of the results is facilitated by the `UV/IR
correspondence' \,\cite{UVIR,Polchinski,HIM3}\,, which can be viewed as a
manifestation of the uncertainty principle in the context of AdS/CFT. To
formulate this correspondence it is preferable to use the inverse radial
coordinate $\chi\equiv \pi R^2/r$, in terms of which the Minkowski
boundary lies at $\chi=0$ and the BH horizon at $\chi=1/T$. Then, the
UV/IR correspondence states that the distance $\chi$ for the penetration
of the wave packet $A_\mu$ in the 5th dimension is proportional to the
transverse size $L$ of the partonic fluctuation of the current in the
physical space (i.e., on the Minkowski boundary): $\chi\propto L$ (see
Figs. 2 and 4 below for some graphical illustrations).

For instance, a {\em space--like} wave in the vacuum ($T=0$) penetrates
in $AdS_5$ up to a maximal distance $\chi_{\rm max}\sim 1/Q$, with
$Q^2\equiv |\omega^2 -q^2|$, and it does so {\em
diffusively}\,\cite{HIM3}\,: the average position $\chi$ of the wave
packet grows like $\chi(t)\sim\sqrt{t/\omega}$ up to a time $t_{\rm
coh}\sim \omega/Q^2$, when $\chi(t_{\rm coh})\sim 1/Q$. This corresponds
to the fact that the physical current fluctuates into a system of partons
(a quark--antiquark pair, or a pair of scalars, or some more complicated
partonic configuration), which diffusively expands in transverse
directions, $L(t)\sim\sqrt{t/\omega}$, up to a maximal size $L_{\rm
max}\sim 1/Q$.

A {\em time--like} current, on the other hand, can decay into the
massless partons of ${\mathcal N}=4$ SYM, so its transverse size can
increase for ever (at least in the vacuum). And indeed the AdS/CFT
calculation\footnote{See also Ref. \cite{Karch} for a similar calculation
involving an open string instead of the Maxwell wave.} \cite{HIM3} shows
that, after the early diffusion at times $t\lesssim t_{\rm coh}$, the
Maxwell wave packet propagates inside $AdS_5$ at a constant speed\,:
$\chi(t)\sim \sqrt{1-v_z^2}\,t$, where $v_z\equiv q/\omega \le 1$. Note
that this radial velocity $v_\chi=\sqrt{1-v_z^2}$ is the same as the
transverse velocity $v_\perp=\sqrt{1-v_z^2}$ of a massless particle with
longitudinal velocity $v_z$. Thus the UV/IR correspondence ($L(t)\sim
\chi(t)\sim \sqrt{1-v_z^2}\,t$) seems to imply that the time--like
current splits into a pair of free, massless, partons which move out with
a common longitudinal velocity $v_z=q/\omega$ (as inherited from the
current) while separating from each other in transverse directions at
velocity $v_\perp=\sqrt{1-v_z^2}$. This would be the correct picture in
lowest--order perturbation theory, but cannot also be the right picture
at strong coupling, where there is no reason why parton branching should
stop at the 2--parton level. Rather, the two partons produced by the
first splitting have a large probability to further radiate before
getting on--shell, and the emitted fields will then radiate in their own,
thus eventually producing some complicated parton configuration.

One can easily check that a simple model assuming `quasi--democratic
branching'\, \cite{HIM3} --- a natural picture at strong coupling, in
which energy and virtuality are almost equally divided among the daughter
particles at each splitting --- leads to the same parametric form for
$L(t)$ as the AdS/CFT calculation. In this picture one has
 \beq\label{BRvacuum}
 \omega_n\,\sim\,\frac{\omega_{n-1}}{2}\,\sim\,\frac{\omega}{2^n}\,,
 \qquad Q_n\,\sim\,\frac{Q_{n-1}}{2}\,\sim\,\frac{Q}{2^n}\,,\qquad
  \Delta t_n\,\sim\,\frac{\omega_{n}}{Q_n^2}\,,\eeq
where $n=0,1,2,\,...$ is the generation index and the lifetime $\Delta
t_n$ of the $n$th parton generation has been estimated via the
uncertainty principle. This implies
 \beq\label{dQdt}
  \frac{Q_n-Q_{n-1}}{\Delta t_n}\, \sim\,-\,\frac{Q}{\omega}\,Q_n^2
 \quad\Longrightarrow\quad \frac{\rmd Q(t)}{\rmd t}\,\simeq\,
 \,-\,\frac{Q^2(t)}{\gamma}\,,\eeq
where $\gamma\equiv \omega/Q = 1/\sqrt{1-v_z^2}$ is the Lorentz factor
for the incoming virtual current, and also for all the other, virtual,
partons produced via successive branching (since the ratio $\omega_n/Q_n
\approx \omega/Q$ is approximately constant during the branching
process). Eq.~(\ref{dQdt}) together with the uncertainty principle
$L(t)\sim 1/Q(t)$ implies that the transverse size of the partonic system
increases like $L(t)\sim \sqrt{1-v_z^2}\,t$, in qualitative agreement
with the respective AdS/CFT result.

\begin{figure}[t]
\centerline{
\includegraphics[width=.9\textwidth]{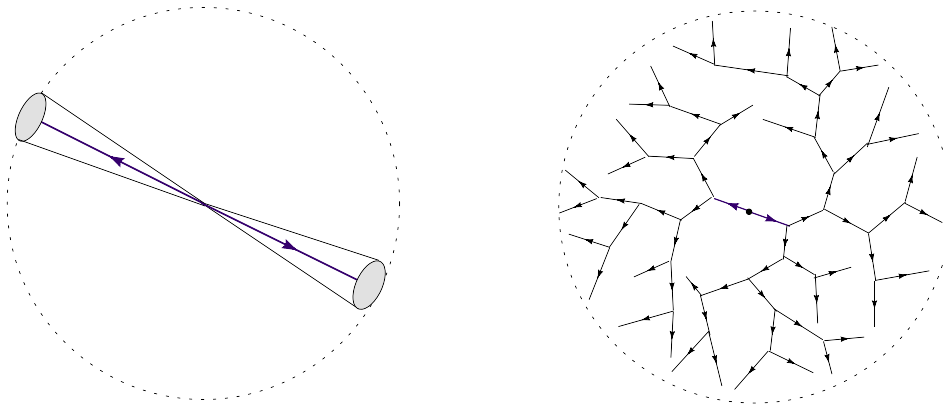}}
\caption{\sl Final state produced in $e^+e^-$ annihilation: (left) weak coupling;
(right) strong coupling.
}
\end{figure}

This picture of `quasi--democratic branching' (that one should think off
as a kind of mean field approximation to the actual dynamics in the gauge
theory at strong coupling) has some consequences for the final state
produced via the decay of a time--like current (so like in $e^+e^-$
annihilation): unlike what happens in QCD at weak coupling, where the
respective final state involves only a few, well collimated, jets --- the
most likely configuration being the 2--jet one illustrated in the l.h.s.
of Fig.~1 ---, at strong coupling the partons will keep fragmenting into
softer and softer partons, until the parton virtuality will degrade down
to a value of the order of the infrared cutoff: $Q_N\sim\Lambda$ for the
last, $N$th, generation. This means that the final state will involve a
large number of particles, $N_{part}\sim 2^N\sim Q/\Lambda$, which will
compose a fully isotropic distribution, as illustrated in the r.h.s. of
Fig.~1. In other terms, there will be no jets in $e^+e^-$ annihilation at
strong coupling, at sharp variance to the corresponding situation in QCD,
as predicted by weak--coupling calculations and actually seen in the
experiments. Similar conclusions have been reached in Refs. \cite{Hofman}
via quite different arguments.

\section{Deep inelastic scattering off a strongly coupled plasma}

We now return to the case of an ${\mathcal R}$--current propagating
through the strongly coupled ${\mathcal N}=4$ SYM plasma, which is our
main physical interest. As we shall see, the corresponding AdS/CFT
results are again suggestive of a `quasi--democratic branching' picture,
which is now generalized to accommodate the effects of the plasma. We
focus on a space--like current which has a large virtuality, $Q\gg T$, as
appropriate for a `hard probe'. It is moreover interesting to choose this
current to have a very high energy (or longitudinal momentum) in the
plasma rest frame, such that $q\sim\omega\gg Q$. This is so since, as we
shall shortly argue, a low energy current does not interact with the
plasma within the present, large--$N_c$, approximation. Via a suitable
change of function, the Maxwell equations for $A_\mu$ can be rewritten as
a pair of time--independent Schr\"odinger equations --- one for the
longitudinal modes, the other one for the transverse ones. Then, the
dynamics can be easily understood by inspection of the respective
potential, as illustrated in Figs. 2 and 4 for two different regimes of
energy. (Note that in plotting the potential in these figures we are
using the dimensionless variables $K\equiv Q/T$ and $k\equiv q/T$; also,
$\chi$ is multiplied by $T$.)

\begin{figure}[t]
\centerline{
\includegraphics[width=6.5cm]{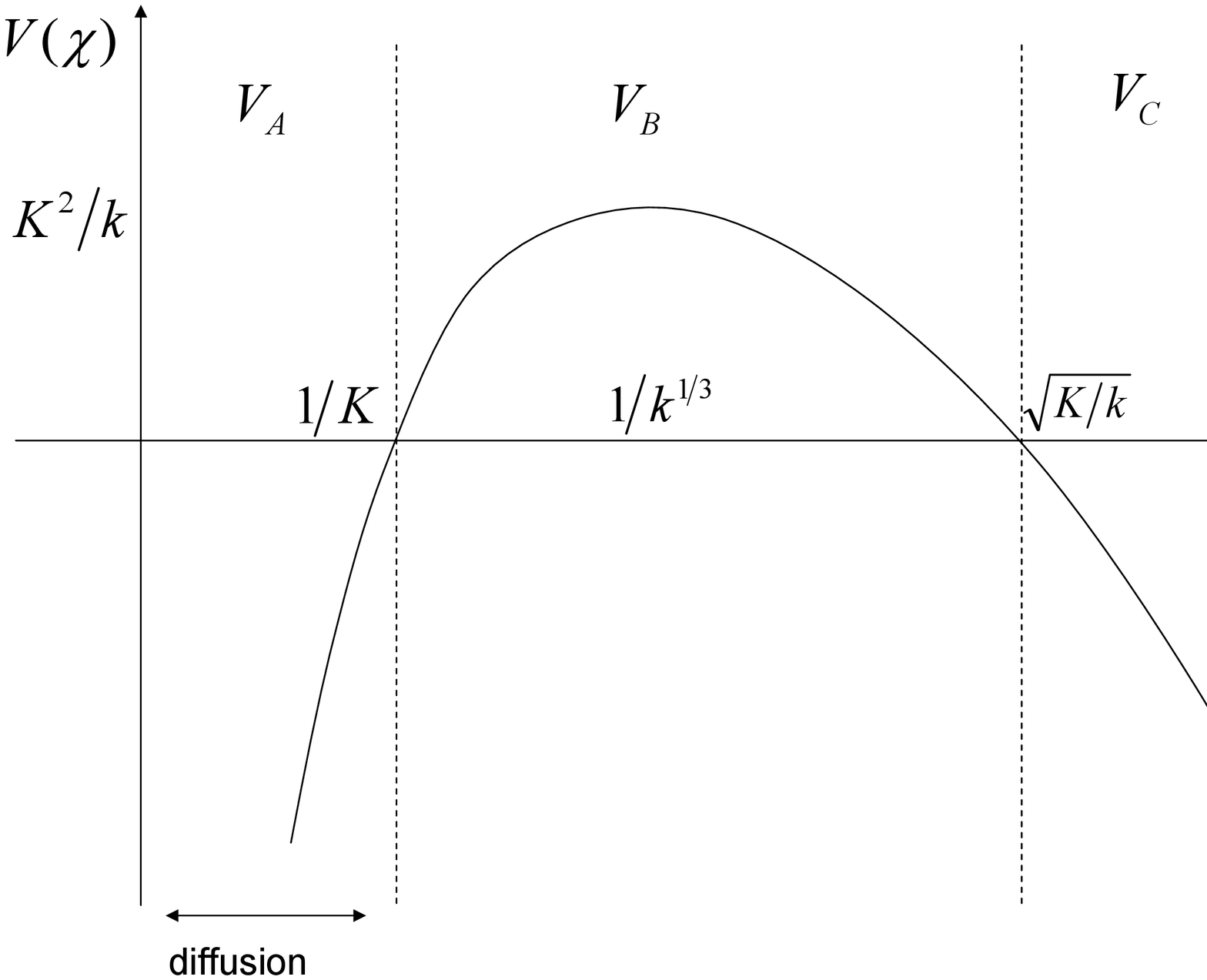}
\hspace*{1.5cm}\includegraphics[width=7.cm]{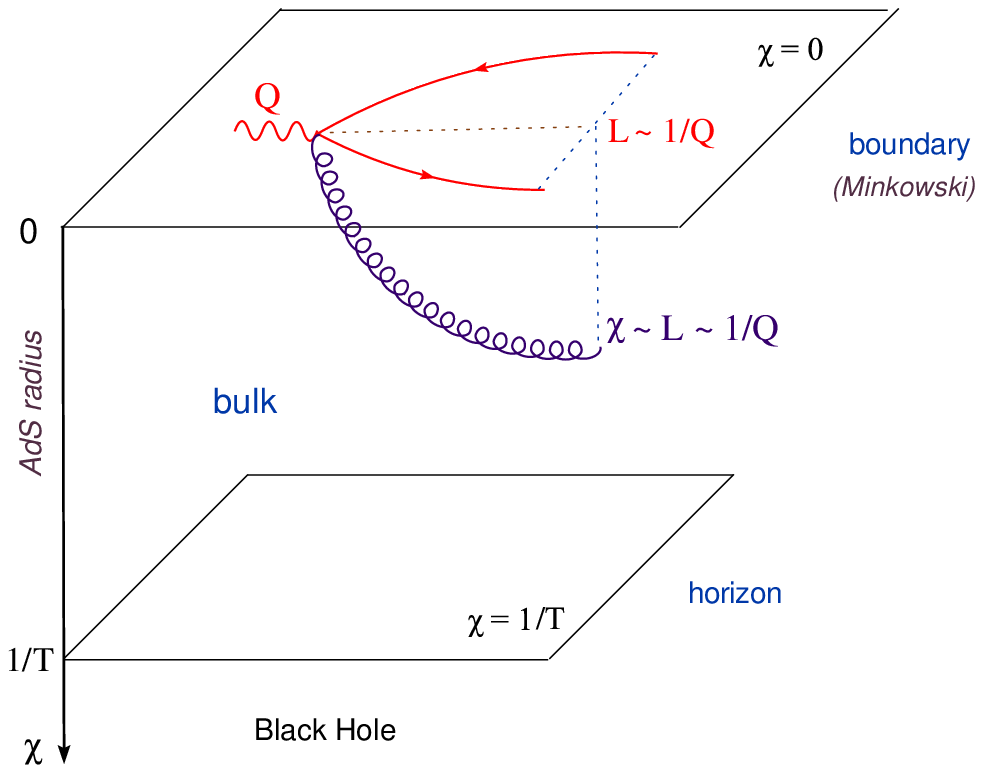}}
\caption{\sl Space--like current in the plasma at not too high energy ($x\gg T/Q$).
Left: the potential barrier. Right: the trajectory of the wave packet
in $AdS_5$ and its `shadow' on the boundary.
}
\end{figure}

Specifically, the dynamics depends upon the competition between, on one
hand, the virtuality $Q^2$, which acts as a potential barrier preventing
the Maxwell wave $A_\mu$ to penetrate deeply inside $AdS_5$, and, on the
other hand, the ratio $\omega^2 T^4/Q^4$, which controls the strength of
the interactions between this wave and the BH at the position $\chi\sim
1/Q$ of the wave packet. (We recall that the gravitational interactions
are proportional to the energy density of the two systems in interaction
--- here, $\omega^2$ for the Maxwell wave and $T^4$ for
the plasma --- and that they show a power fall--off at large separations
--- here, they fall with $r$ like $1/r^4$, with $r\sim 1/\chi\sim Q$.)
The two forces in action equilibrate each other (parametrically) when
 \beq\label{force} Q\,\sim\,\frac{\omega T^2}{Q^2}\,,
 \eeq
a condition which suggests the following physical interpretation back in
the gauge theory: the r.h.s. of Eq.~(\ref{force}) can be rewritten as the
product $t_{\rm coh}\times T^2$ between the lifetime $t_{\rm coh}\sim
\omega/Q^2$ of a partonic fluctuation and a quantity ($T^2$) which has
the dimension of a force. This implies that, at strong coupling and large
$N_c$, the plasma acts on the colored partons with a constant force $\sim
T^2$. A closer inspection\cite{HIM3} shows that this force acts towards
reducing the parton virtuality or, equivalently, increasing their
transverse size
 : $\rmd Q/\rmd t\sim - T^2$. Then Eq.~(\ref{force}) can be
recognized as the condition that the mechanical work done by this force
during the parton lifetime be large enough to compensate the potential
barrier due to the virtuality.

We are therefore led to distinguish between two physical regimes,
according to the value of the dimensionless parameter $Q^3/\omega T^2$
(which becomes of order 1 when the condition (\ref{force}) is satisfied).
This parameter can be conveniently rewritten as $xQ/T$, where $x\equiv
Q^2/2\omega T$ is the Bjorken variable for DIS off the plasma and has the
meaning of the longitudinal momentum fraction of the `parton' struck by
the current (in the plasma infinite momentum frame).

\texttt{(i)} In the high--$Q^2$ (or large--$x$) regime at $Q^3/\omega T^2
\gg 1$ (or $x\gg T/Q$), the interaction with the plasma is relatively
weak and the dynamics is almost the same as in the vacuum: the wave
penetrates in $AdS_5$ up to a maximal distance $\chi\sim 1/Q$ where it
gets stuck against the potential barrier (see Fig. 2). Physically, this
means that the current fluctuates into a system of partons with
transverse size $L\sim 1/Q$ which is essentially insensitive to the
plasma. At finite temperature, however, the potential barrier has only a
finite width --- it extends up to a finite distance $\chi\sim
(1/T)\sqrt{Q/\omega}$ ---, so there is a small, but non--zero,
probability for the wave to cross the barrier via tunnel effect.
Physically, this means that the plasma structure function at large $x$ is
non--vanishing, but extremely small (exponentially suppressed):
$F_{2}(x,Q^2)\,\propto\,{xN^2_cQ^2}\, \exp\{-(x/x_s)^{1/2}\}$ for $x \gg
x_s\equiv T/Q$. Therefore, when probing the plasma on a transverse
resolution scale $Q^2$, one finds that there are essentially no partons
with momentum fraction $x$ larger than $x_s=T/Q\ll 1$.

\begin{figure}[t]
\centerline{
\includegraphics[width=0.9\textwidth]{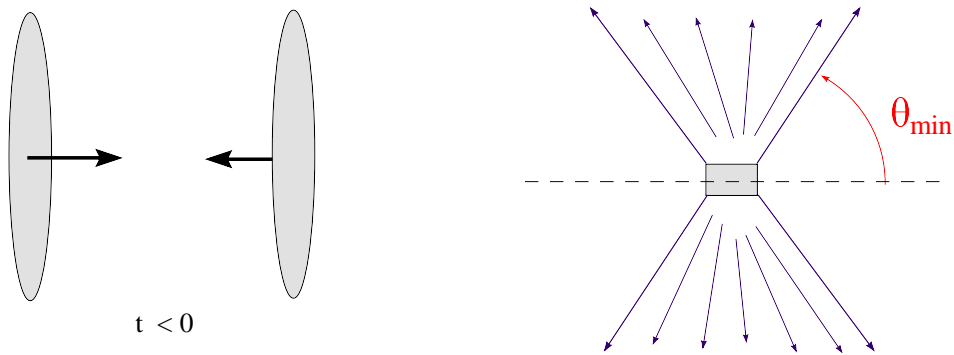}}
\caption{\sl A picture of a hypothetical hadron--hadron collision at strong
coupling: there is no particle production at either forward, or backward,
rapidities.
}
\end{figure}

A similar conclusion holds for the case where the target is a
dilaton\cite{Polchinski,HIM1} instead of a plasma; in that case one
finds\cite{HIM1} $x_s\sim \Lambda^2/N_c^2Q^2$, which is extremely small
when $N_c\gg 1$ and $Q\gg\Lambda$. This has dramatic consequences for a
(hypothetical) hadron--hadron collision at strong coupling, in which the
partons from the incoming wave functions would be liberated: since there
are no partons carrying large fractions of the longitudinal momenta of
the original hadrons, there will no particles produced at either forward,
or backward, rapidities (see Fig.~3). Rather, particle production would
be limited to central rapidities alone.

\texttt{(ii)} What happened to the partons then ? To answer this
question, let us explore smaller values of Bjorken's $x$, say, by
increasing the energy $\omega$ at fixed $Q^2$ and $T$. Then the barrier
shrinks and eventually disappears, when the energy is such that the
condition (\ref{force}) is satisfied. This condition can be solved for
the virtuality $Q$, in which case it yields the {\em plasma saturation
momentum}  $Q_s^2(x,T)\sim T^2/x^2$, or, alternatively, for the Bjorken
$x$ variable, thus yielding $x_s(Q,T)\sim T/Q$. For even higher energies,
meaning $x < x_s$, the barrier has disappeared and the Maxwell wave can
propagate all the way down to the black hole, into which it eventually
falls, along a trajectory which coincides with the `trailing string'
characterizing the energy loss of a heavy quark \cite{Drag} (see Fig.~4).
Physically, this means that the current has completely dissipated into
the plasma.

\begin{figure}[tb]
\centerline{
\includegraphics[width=6.5cm]{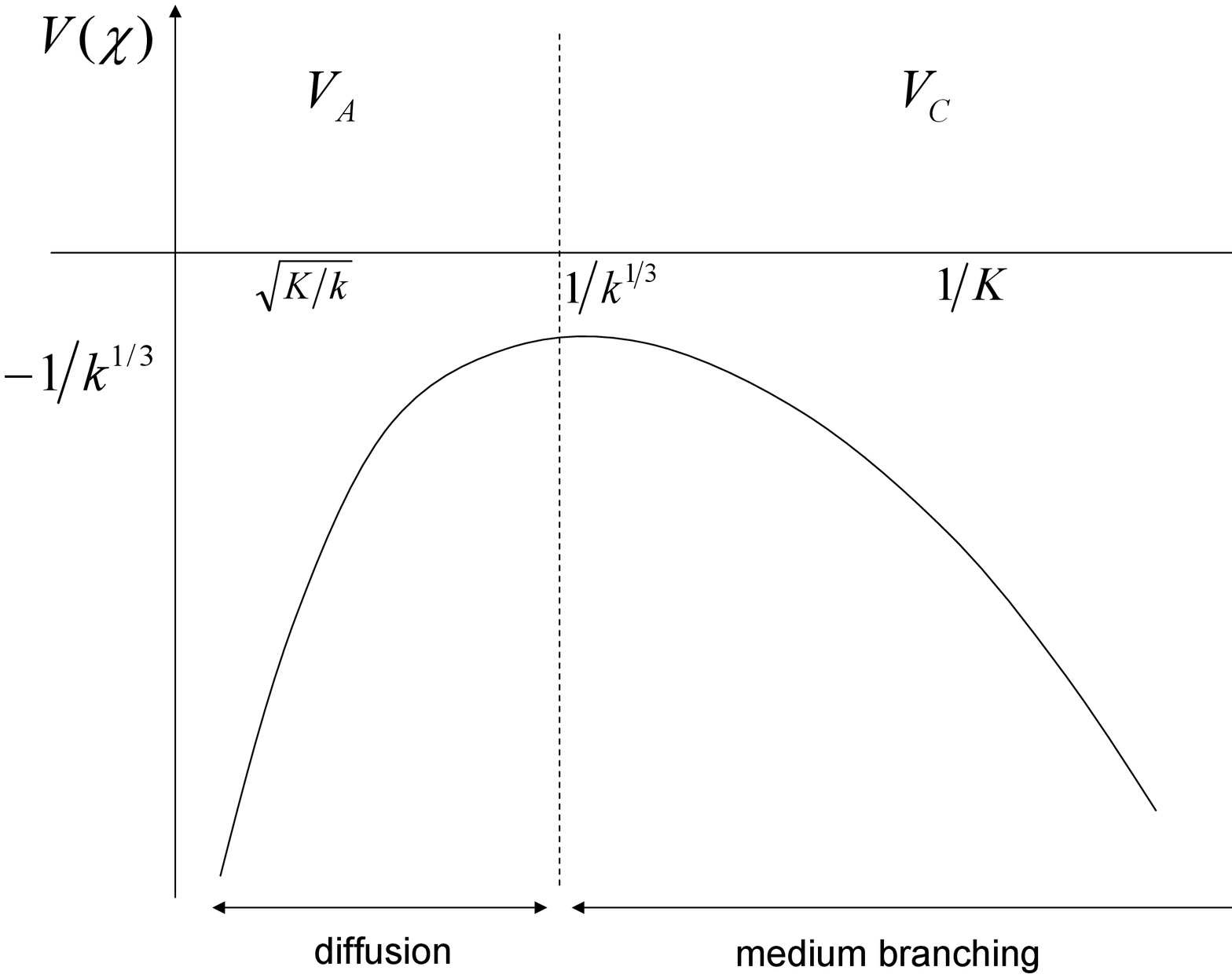}\hspace*{1.5cm}
\includegraphics[width=7.cm]{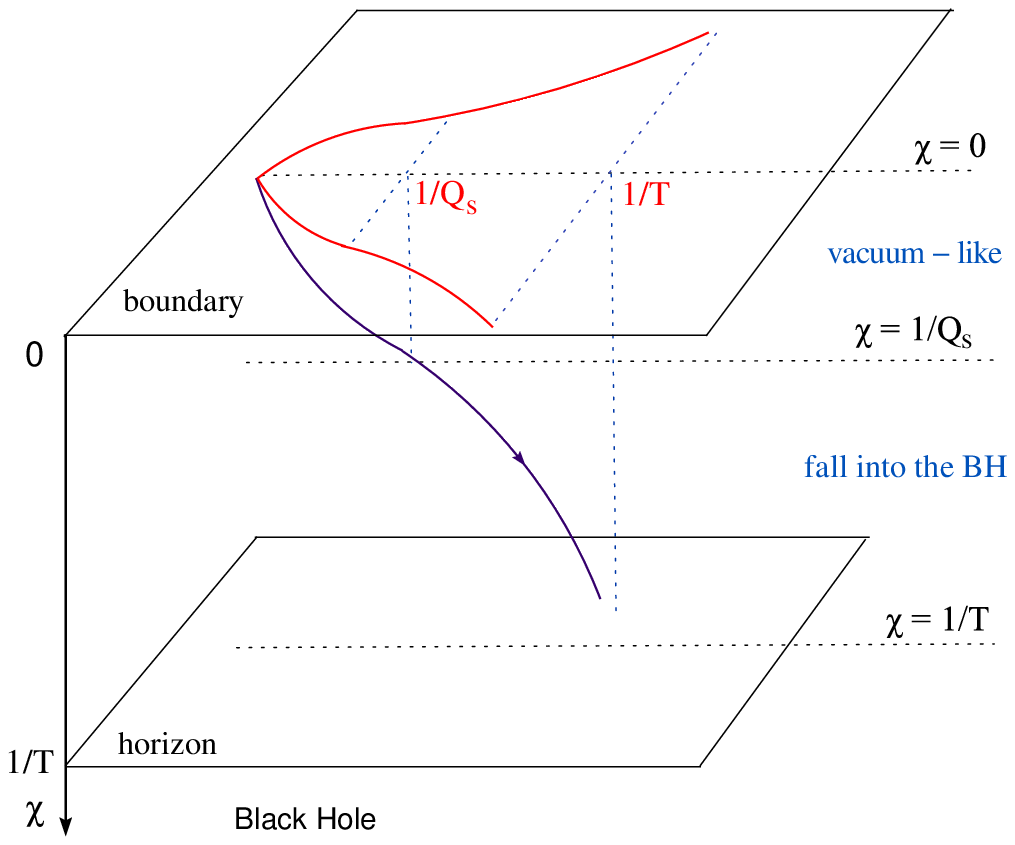}}
\caption{\sl High--energy current in the plasma ($x\lesssim T/Q$).
Left: the potential. Right: the fall of the wave packet
into the black hole and its `shadow' on the boundary.}
\end{figure}

From the point of view of DIS, this situation corresponds to the
unitarity, or `black disk', limit, i.e., the strongest possible
scattering. Hence, for $x < x_s(Q)$, the plasma structure function
$F_2(x,Q^2)$ is non--vanishing and large. Specifically, the AdS/CFT
calculation yields\cite{HIM2} $F_{2}(x,Q^2)\,\sim\,{xN^2_cQ^2}$ for
$x\lesssim x_s$, a result which admits a natural physical interpretation:
for a given resolution $Q^2$, essentially all partons have momentum
fractions $x\lesssim T/Q\ll 1$ and occupation numbers
$n\sim\mathcal{O}(1)$. Alternatively, for a given value of $x\ll 1$,
partons exist only at sufficiently low transverse momenta
$k_\perp\lesssim Q_s(x)$. This is similar to parton saturation in pQCD,
except that, now, the occupation numbers at saturation are of order one,
rather than being large ($n\sim 1/g^2 N_c$), as it is the case at weak
coupling\cite{CGC}\,.

By using the above result for $F_2$, one can check that the energy
sum--rule is satisfied, as it should --- the total energy density in the
plasma, as probed on a resolution scale $Q^2$, is independent of $Q^2$
and of order $N_c^2T^4$\,:
 \beq
 \mathcal{E}\,= \,T^2\int_0^1\rmd x\,F_2(x,Q^2)\sim
 T^2\,\Big[xF_2(x,Q^2)\Big]_{x=x_s}\,\sim\,N_c^2T^4 \,.
 \eeq
Moreover, as emphasized in the above estimate for $\mathcal{E}$, this
sum--rule is dominated by $x\sim x_s$\,: in the plasma infinite momentum
frame, all of the plasma energy is carried by partons lying along the
saturation line $x=x_s(Q)$.

Note that the plasma saturation momentum appears to rise much faster with
$1/x$ then the corresponding scale for a `dilaton' \cite{HIM2} (at strong
coupling in both cases): we have indeed $Q_s^2\sim 1/x^2$ for the plasma,
as opposed to $Q_s^2\sim 1/x$ for a single dilaton target. The additional
factor of $1/x$ in the plasma case is however rather trivial, as it
merely reflects the kinematics: it comes from the fact that the
interaction length for the current in the plasma is set by its coherence
time $t_{\rm coh}\sim \omega/Q^2\sim 1/(xT)$, and thus it increases like
$1/x$.

Since the current is tantamount to a `meson' with size $1/Q$ and rapidity
$\gamma=\omega/Q$, the above discussion implies an upper limit on the
transverse size of this `meson' before it melts in the plasma: $L_{\rm
max}\sim 1/Q_s \sim 1/\sqrt{\gamma}\,T$. This limit is consistent with
the meson screening length computed in Refs. \cite{Meson}\,. The lifetime
of the current (estimated as the duration of the fall into the BH) is
found as $\Delta t\sim \omega/Q_s^2 \propto \omega^{1/3}$, in agreement
with a recent estimate of the `gluon' lifetime in Ref.
\cite{GubserGluon}\,.

This similarity between seemingly different calculations which refer to
various types of projectile (heavy quarks, massless gluons, of the
${\mathcal R}$--current) suggests that the mechanism responsible for
energy dissipation in the plasma must be universal and that it acts at
partonic level. Our previous discussion of `quasi--democratic' parton
branching in the vacuum (cf. Sect.~3) together with the observation that
the plasma acts on partons with a constant force $\rmd Q/\rmd t\sim -
T^2$ makes it natural to interpret the dissipation in the plasma as the
result of {\it medium--induced branching} : the current fragments into
partons via successive branchings, with a splitting rate which is
amplified by the temperature (compare to Eq.~(\ref{BRvacuum})) :
 \beq\label{BRplasma}
 \omega_n\,\sim\,\frac{\omega_{n-1}}{2}\,,
 \qquad  \Delta t_n\,\sim\,\frac{\omega_{n}}{Q_n^2}
 \,,\qquad \frac{\Delta Q_n}{\Delta
t_n}\, \sim\,-\,T^2\,\Longrightarrow \, Q_n \,\sim\,(\omega_nT^2)^{1/3}
 \,.\eeq
The last estimate for $Q_n$ (the virtuality of a parton in the $n$th
generation) in the above equation follows by assuming that $Q_n$ and
$Q_{n-1}$ are of the same order of magnitude and then using the shown
expressions for $\Delta t_n$ and for $\Delta Q_n$. This kind of
medium--induced branching starts operating when the current reaches a
critical size $L_{\rm max}\sim 1/Q_s$ (i.e., when the dual Maxwell wave
reaches the `point of no return' $\chi_s\sim 1/Q_s$, cf. Fig.~4), and
then it continues until the energy and the virtuality of the partons
degrade down to values of order $T$. As it can be easily checked on
Eq.~(\ref{BRplasma}), this whole process takes a time $\Delta t\sim
Q_s/T^2\sim \omega/Q_s^2$, which is much shorter --- in this high--energy
regime where $Q^2_s\gg Q^2$ --- than the `coherence' time $t_{\rm
coh}\sim \omega/Q^2$ required for the formation of a nearly on--shell
partonic fluctuation. Hence, the high--energy current looses energy very
fast and disappears into the plasma long before having the time to
fluctuate into nearly on--shell partons.

As detailed in Refs. \cite{HIM2,HIM3,QSAT}\,, this branching scenario is
consistent with all the relevant results emerging from AdS/CFT
calculations. For instance, the enveloping curve of the partonic system
produced through branching (before the partons disappear into the plasma)
coincides with the `trailing string' solution of Refs. \cite{Drag}, as it
should by virtue of the UV/IR correspondence. Furthermore, this scenario
correctly reproduces the parametric forms of the laws describing the
energy loss and the transverse momentum broadening of a heavy quark, as
originally computed from AdS/CFT in Refs. \cite{Drag,Broad}\,. For a
ultrarelativistic quark ($\gamma\gg 1$) one finds\cite{HIM2,QSAT}
 \beq\label{dragf}
 -\frac{\rmd E}{\rmd t}\,\sim\,\sqrt{\lambda}\,Q_s^2\,,\qquad
 \frac{\rmd \langle p_\perp^2\rangle}{\rmd
 t}\,\sim\,\sqrt{\lambda}\,T^2 Q_s\,,\eeq
for the rate of change in the energy and in the average transverse
momentum squared, respectively. In these equations, $Q_s$ is the plasma
saturation momentum corresponding to the maximum--energy partons which
are freed into the plasma (i.e., those partons whose emission gives the
largest contributions to the change in energy and momentum). These
partons have $q_\perp\sim Q_s$ and $\omega\sim\gamma Q_s$. By using these
estimates and recalling that $Q_s^2\sim \omega T^2/Q_s \sim \gamma T^2$,
one can easily check that Eqs.~(\ref{dragf}) are indeed consistent with
the respective results in Refs. \cite{Drag,Broad}\,. The derivation of
these equations from the medium--induced branching
scenario\cite{HIM2,QSAT} clarifies their physical interpretation, thus
emphasizing the fact that the mechanism for momentum broadening at strong
coupling is very different from the corresponding mechanism at weak
coupling\cite{BDMPS} : namely, this proceeds via medium--induced parton
radiation in the former case, as opposed to thermal rescattering in the
latter.

To summarize, the picture of a plasma as revealed by hard probes and,
more generally, the overall picture of high--energy scattering appear to
be quite at strong coupling as compared to what we expect in pQCD and we
actually see in experiments: at strong coupling there are no jets in
$e^+e^-$ annihilation, no forward/backward particle production in
hadron--hadron collisions, no partons except at very small $x$, and there
are different physical mechanisms at work, which control the jet energy
loss and its transverse momentum broadening. Such differences suggest
that much caution should be taken when trying to extrapolate results from
AdS/CFT to QCD in the particular context of the hard probes. A possible
strategy in that sense, as recently suggested in \cite{Mueller08}\,, is
to distinguish between `hard' and `soft' momentum contributions to the
observables measured by hard probes, say by introducing a `semi--hard'
separation scale $Q_0$, and then use AdS/CFT techniques in the soft
sector ($k_\perp\le Q_0$) alone --- the hard sector being treated in
perturbation theory.

\comment{
\subsection*{Acknowledgments}
I would like to thank the organizers of the Yukawa International Program
for Quark--Hadron Sciences (YIPQS) {\em New Frontiers in QCD 2008}, hold
at Yukawa Institute for Theoretical Physics, for hospitality and support
during my participation in this program.
 }


\end{document}